\documentclass[11pt]{article}

\usepackage{amssymb}
\usepackage{amsthm}
\usepackage{amsmath}

\usepackage[mathscr]{eucal}
\usepackage{graphicx} 
\usepackage{psfrag}
\usepackage{subfigure}

\usepackage{fancyhdr}

\graphicspath{{.}{/afs/aei-potsdam.mpg.de/u/jmh/figures/drawn_by_xfig/BianchiIVlasov/}}

\renewcommand{\sectionmark}[1]%
        {\markboth%
                {}%
                {{\rm\thesection}\quad{\sc #1}}}

\theoremstyle{plain}
\newtheorem{theorem}{Theorem}[section]
\newtheorem{corollary}[theorem]{Corollary}
\newtheorem{lemma}[theorem]{Lemma}

\theoremstyle{remark}

\newtheorem*{example}{Example}
\newtheorem*{remark}{Remark}

\begin{document}

\title{\LARGE\scshape Bounds on $\mathrm{2m/r}$ for static perfect fluids} 

\author{ \\
{\large\sc J.\ Mark Heinzle}\thanks{Electronic address:  {\tt Mark.Heinzle@univie.ac.at}} \\[1ex]
Gravitational Physics, \\
Faculty of Physics, University of Vienna, \\
A-1090 Vienna, Austria}

\date{\large{\today}}
\maketitle
\begin{abstract}

For spherically symmetric relativistic perfect fluid models,
the well-known Buchdahl inequality provides the bound $2 M/R \leqslant 8/9$,
where $R$ denotes the surface radius and $M$ the total mass of a solution.
By assuming that the ratio $p/\rho$ be bounded, where $p$ is the pressure,
$\rho$ the density of solutions, we prove a sharper
inequality of the same type, which depends on the actual bound imposed on $p/\rho$. 
As a special case, when we assume the
dominant energy condition $p/\rho \leqslant 1$, we obtain $2 M/R \leqslant 6/7$.

\end{abstract}

\vfill
\newpage

\section{Introduction}
\label{introduction}

In General Relativity, the masses $M$ of equilibrium configurations with a
given fixed radius $R$ are limited.
Indeed, from the theory of black holes,
it is clear that $2 M/R < 1$ is required,
since otherwise apparent horizons
must necessarily appear.
In 1959, H.A.~Buchdahl~\cite{Buchdahl:1959} 
proved a stricter inequality:
\begin{equation}\label{buchfirst}
\frac{2 M}{R} \leq \frac{8}{9}
\end{equation}
Buchdahl's inequality holds for spherical stellar models whose matter
is ordinary perfect fluid matter satisfying a non-decreasing
barotropic equation of state $\rho= \rho(p)$ relating the (positive) 
energy density $\rho$ and the (positive) pressure $p$. 
The major appeal of Buchdahl's inequality lies in the fact 
that it leaves a gap between the maximally attainable value of $2 M/R$ and
the threshold of horizon formation.
Buchdahl's inequality is a classical result in General Relativity
and a discussion of it is included in almost every textbook.
Its significance in astrophysics is due to the fact that~\eqref{buchfirst}
yields a bound on the surface red shift (gravitational red shift)
of stellar models.

There exists several 
bounds on $2 M/R$ under different assumptions than Buchdahl's.
In~\cite{Guven/OMurchadha:1999}, the inequality~\eqref{buchfirst}
is shown to hold also for anisotropic matter, where
the tangential pressure is assumed to be less than the radial pressure.
In~\cite{Andreasson:2007}, collisionless matter is considered
instead of fluid matter; in addition, a very general
inequality is derived that is independent of the specifics of
the matter content. 
Recently, the authors of~\cite{Karageorgis/Stalker:2007}
considered a large variety of assumptions.
In particular, they obtain inequalities by imposing
merely the dominant energy condition; see also the remarks in the conclusions.

In this paper we return to Buchdahl's original realm, i.e.,
we investigate ordinary perfect fluid matter that satisfies a non-decreasing 
equation of state.
Although the Buchdahl inequality~\eqref{buchfirst} is sharp under
these assumptions, the solution that maximizes the quantity $2 M/R$
is rather unphysical: The stellar model satisfying $2 M/R = 8/9$
consists of an incompressible fluid, i.e., $\rho = \mathrm{const}$
throughout the entire configuration, and $p$ diverges
as $r\rightarrow 0$. Accordingly, for this limiting solution, 
$p/\rho$ is unbounded, so that
in particular the dominant energy condition is violated.
In this paper we derive a generalization of
Buchdahl's inequality that applies if---in addition to
Buchdahl's original assumptions---we also assume a 
bound on $p/\rho$; this new inequality
is formulated in Theorem~\ref{thethm}.
Its major property can be condensed into the following key
statement: The more restrictive the bound on $p/\rho$, 
the more restrictive the bound on $2 M/R$. 
The extremal cases covered by Theorem~\ref{thethm} are
the case $p/\rho \leqslant \infty$ (i.e., no bound), where the original
Buchdahl inequality~\eqref{buchfirst} is recovered,
and the case where $p/\rho \leqslant \varepsilon$ for
an arbitrarily small $\varepsilon>0$ (which is the most restrictive bound), which leads to 
$2 M/R \leqslant \mathrm{const}\:\varepsilon$.
When we assume the dominant energy condition (i.e., the bound $p/\rho <1$), 
we find that
\begin{equation}
\frac{2 M}{R} \leq \frac{6}{7}\:.
\end{equation}
Since the methods of our proof are based on the dynamical systems approach
to relativistic perfect fluid models, which is due to~\cite{Heinzle/Rohr/Uggla:2003},
we begin by briefly introducing these techniques.

\section{Dynamical systems formulation}
\label{dynamicalsystemsformulation}

Static spherically symmetric 
relativistic perfect fluid solutions (relativistic stellar models) 
are solutions of 
the relativistic equations of hydrostatic equilibrium:
\begin{subequations}\label{oppvol}
\begin{align}
\label{mrov}
  \frac{dm}{dr} &= 4\pi r^2 \rho\, , \\
\label{oppvoleq}
  \frac{dp}{dr} &= -\frac{Gm\rho}{r^2}\left(1 + \frac{p}{\rho c^2}\right)
  \left(1 + \frac{4\pi r^3 p}{m c^2}\right)
  \left(1 - \frac{2G m}{c^2 r}\right)^{-1}\,.
\end{align}
\end{subequations}
Equation~\eqref{oppvoleq} is the famous
Tolman-Oppenheimer-Volkoff (TOV) equation~\cite{Oppenheimer/Volkoff:1939, Tolman:1939}.
The function $m = m(r)$ is the so-called mass function; when $R$ is the surface
radius of a stellar model, we denote by $M= m(R)$ its total mass.
A solution of~\eqref{oppvol} gives rise to a static spherically symmetric
perfect fluid spacetime, which is represented by the metric
\begin{equation}\label{ds2}
ds^2= -c^2 e^{2\nu(r)} dt^2 +  \Big(1 - \frac{2G m(r)}{c^2 r}\Big)^{-1}
dr^2 + r^2 \big(d{\theta}^2 + {\sin}^2{\theta}\, d{\phi}^2\big)\, ,
\end{equation}
where the ``relativistic potential'' $\nu$ is given by $d\nu/dr
= -(dp/dr)/(p+\rho c^2)$. Note that $r$ 
is the standard areal radial coordinate. 
To form a determined system, these
equations are supplemented by a barotropic equation of state
$\rho(p)$.

We assume that $\rho$ and $p$ are non-negative and related by
a barotropic equation of state $\rho(p)$ that is
sufficiently smooth for $p>0$. 
(In principle, equations of state that are only piecewise
smooth do not pose any further difficulties; 
however, for simplicity, we confine ourselves to the simplest case.)
In the following we choose units so that $G = 1$ and $c =1$.

Although the system~\eqref{oppvol} is adequate for several purposes,
it is not suitable for probing generic features of the 
solution space associated with large classes of equations of state.
For example, in the standard proof of Buchdahl's inequality, 
see~\cite{Buchdahl:1959} or~\cite{Beig/Schmidt:2000},
the system is replaced by an alternative system that is based
on a choice of variables different from $(m,p,r)$.
To remedy the defects of the TOV equation, 
in~\cite{Heinzle/Rohr/Uggla:2003}, the system~\eqref{oppvol}
is reformulated as a dynamical system. This reformulation
turns out to be the key to the derivation of several
qualitative results
such as theorems on the finiteness of $R$ and $M$, and
results on qualitative features of the $M$-$R$-diagram.
In the following we shall present a brief summary of
this dynamical systems formulation, since it constitutes
the basic ingredient in the proof of our main theorem.

The two-dimensional system~\eqref{oppvol} 
allows a reformulation in terms of a
regular autonomous three-dimensional system of
differential equations defined on a compact state space;
see~\cite{Heinzle/Rohr/Uggla:2003}.
To obtain this dynamical system formulation one
makes a variable
transformation from $(m,p>0,r>0)$ to the 
variables
\begin{equation}\label{uvom}
u = \frac{4\pi r^3 \rho}{m}\ , \quad v = \frac{\rho}{p}
\left(\frac{\frac{m}{r}}{1 - \frac{2 m}{r}}\right)\:,
\quad \omega = \omega(p) = p\ ;
\end{equation}
in principle, $\omega(p)$ can be a more or less arbitrary 
function that is strictly monotonically increasing and
sufficiently smooth. For our purposes, however, the choice 
$\omega = p$ is sufficient.

Since the variables $(u,v,\omega)$ are unbounded, 
we replace these variables by bounded variables $(U,V,\Omega) \in (0,1)^3$
by simply defining
\begin{equation}\label{boundedvar}
U = \frac{u}{1+u}\:,
\qquad V = \frac{v}{1+v} \:,
\qquad \Omega = \frac{\omega}{1+\omega} \ .
\end{equation}
In addition, we introduce a new independent variable $\lambda$,
which is a measure of $r$, by
\begin{equation}
\frac{d\lambda}{d\log r} =  (1 - U)^{-1}(1 - V)^{-1}\:.
\end{equation}
Using these definitions turns~\eqref{oppvol} into
a dynamical system on the bounded state space 
$\mathbf{X} = (0,1)^3 = \{(U,V,\Omega)\,|\, U,V,\Omega \in (0,1)\}$,
which is the unit cube.
The dynamical system reads
\begin{subequations}\label{UVOmega}
\begin{align}
\label{Ueq}
&\frac{dU}{d\lambda} = U(1-U)[(1-V)(3-4U) - \Upsilon\,H] \\
\label{Veq}
&\frac{dV}{d\lambda} = V(1-V)[(2U-1)(1-V+2\sigma\,V) +
(1-\Upsilon)\,H] \\
\label{Omegaeq}
&\frac{d\Omega}{d\lambda} = -\Omega(1-\Omega)H\, , \\[2ex]
& \mbox{where}\quad
 H = H(U,V,\Omega) = (1 + \sigma)V(1 - U + \sigma U)\: .
\end{align}
\end{subequations}
The r.h.~side contains two functions that encode the information
on the equation of state:
\begin{equation}\label{indexdefs}
\Upsilon = \frac{p}{\rho}\, \frac{d\rho}{dp}
\:\,,\qquad\qquad
\sigma = \frac{p}{\rho}\ .
\end{equation}
In the context of the system~\eqref{UVOmega}, these functions
are to be viewed as functions of $\Omega$ (instead of functions of $p$).
For details on the derivation of the equations~\eqref{UVOmega} we refer to~\cite{Heinzle/Rohr/Uggla:2003}.

Given an equation of state, the TOV equation admits
a one-parameter family of \textit{regular solutions}, which are
solutions of~\eqref{oppvol} with a regular center of
spherical symmetry. The parameter characterizing this family
is most conveniently chosen to be the central pressure of the 
solution (i.e., $p_c = p|_{r= 0}$). (For a proof
see~\cite{Rendall/Schmidt:1991}; the
dynamical systems analysis of~\cite{Heinzle/Rohr/Uggla:2003} provides
a simple alternative.)
Since $m \sim 4 \pi r^3\rho_c/3$ in the limit $r\rightarrow 0$
(where $\rho_c = \rho(p_c)$ is the central density),
we find that regular solutions satisfy
\begin{equation}
U \rightarrow \frac{3}{4}\:, \qquad
V \rightarrow 0 \:,\qquad
\Omega \rightarrow \Omega_c
\end{equation}
as $r\rightarrow 0$;
in addition,
\begin{equation}
\lambda \rightarrow -\infty \quad\Leftrightarrow\quad r\rightarrow 0\:.
\end{equation}
Therefore, in the dynamical systems picture, a regular
perfect fluid solution is represented by 
an orbit $\big(U(\lambda),V(\lambda),\Omega(\lambda)\big)$ in the state space $\mathbf{X}$
that emerges from what we call
the line of regular initial values
$\{(U,V,\Omega)\,|\, U=3/4, V= 0, \Omega \in(0,1)\}$.
We conclude by noting that the limit $\lambda\rightarrow \infty$ 
corresponds to the limit $r\rightarrow R$, where $R$ is the
surface radius of the perfect fluid solution; this is
because $\Omega \rightarrow 0$ (and thus $p\rightarrow 0$)
is this limit; for details, see~\cite{Heinzle/Rohr/Uggla:2003}.

\begin{remark}
In the definition of the variable $V$, the
factor $1 - 2 m/r$ appears. This does not pose any problems, 
since this factor 
is always positive for relativistic perfect fluids;
cf.~\cite{Baumgarte/Rendall:1993}.
One can even show that it is not necessary to resort to additional considerations
to ensure that $1 - 2 m/r > 0$; this issue can be dealt with
completely within the dynamical systems formulation:
If $1 - 2 m/r > 0$ holds initially (as is true for regular solutions),
then $1 - 2 m/r > 0$ for all $r$ corresponding to $\lambda \in (-\infty,\infty)$
and thus in particular for all $r < R$. This is simply because $1-2m/r$ can be
written as 
$1 - 2 m/r = (1-V)(1-V+2 \sigma V)$, which is positive, 
since $0< V < 1$ for all $\lambda  \in (-\infty,\infty)$.
To show that also $1 - 2 M/R > 0$ (which corresponds to the limit $r\rightarrow R$,
or, equivalently, to $\lambda\rightarrow \infty$),
a straightforward dynamical systems analysis of the $\omega$-limits
of solutions in $\mathbf{X}$ is sufficient; see~\cite{Heinzle/Rohr/Uggla:2003}.
\end{remark}

\section{A generalization of the Buchdahl inequality}
\label{sec:genbuch}

Consider the function
\begin{equation}\label{Phidef}
\Phi = \Phi(\mathcal{M},\mathcal{P}) =
\frac{9 \mathcal{M} \left[ 2(2 \mathcal{M}+ \mathcal{P}) - (3\mathcal{M}+\mathcal{P})^2\right]}{(1-2\mathcal{M})(3\mathcal{M}+\mathcal{P})^2} \:,
\end{equation}
where
\begin{equation}
\mathcal{M} =   \frac{m}{r} \:,\qquad
\mathcal{P} =  4\pi r^2 p \:.
\end{equation}
For each constant 
\begin{equation}
g \in [ 0, 4)\:,
\end{equation}
the inequality 
\begin{equation}\label{Buchtype}
\Phi \geqslant g
\end{equation}
defines a Buchdahl type inequality. This is simply because 
it is an inequality relating $m/r$ with the pressure $p$.

\begin{example}
In the special case $g = 0$, the inequality~\eqref{Buchtype} reduces to
\begin{equation}\label{Buchdahlclassic}
2(2 \mathcal{M}+ \mathcal{P}) - (3\mathcal{M}+\mathcal{P})^2 \geqslant 0 \:;
\end{equation}
alternatively, when we use standard variables, we obtain 
\begin{equation}\label{Buchdahlclassclass}
\frac{2 m}{r} \leqslant \frac{4}{9} \left( 1 - 6 \pi r^2 p + \sqrt{1 + 6 \pi r^2 p} \right)\:,
\end{equation}
which is the (complete) Buchdahl inequality.
Consider a regular perfect fluid solution $(m(r),p(r))$ that satisfies this inequality; 
let $R$ denote the surface radius, which is determined by the vanishing of the pressure,
and $M = m(R)$ the total mass of the solution.
Setting $\mathcal{P} = 0$ in~\eqref{Buchdahlclassic},
or $p=0$ in~\eqref{Buchdahlclassclass}, which corresponds to evaluating the inequality
at the surface $R$, yields the so-called Buchdahl surface inequality
\begin{equation}
\frac{2 M}{R} \leqslant \frac{8}{9}\:.
\end{equation}
The Buchdahl inequality is due to H.~A.~Buchdahl~\cite{Buchdahl:1959};
it holds for all regular perfect fluid solutions
that obey an equation of state that is non-decreasing.
\end{example}

Setting $\mathcal{P} = 0$ in~\eqref{Buchtype}
(for an arbitrary value of $g \in [0,4)$) corresponds to
evaluating this inequality at the surface of a perfect fluid solution.
A simple algebraic manipulation shows that $\Phi|_{\mathcal{P} = 0} \geqslant g$
is equivalent to $\mathcal{M}|_{R} \leqslant (4 - g)/(9 - 2 g)$ or
\begin{equation}\label{Buchtypesurface}
\frac{2 M}{R} \leqslant \frac{8 - 2 g}{9-2 g}\:.
\end{equation}
The inequality~\eqref{Buchtype} thus naturally incorporates a surface inequality
of the type~\eqref{Buchtypesurface}.

\begin{theorem}\label{thethm}
Consider an equation of state $\rho = \rho(p)$ that is non-decreasing; let $\sigma = p/\rho$.
Every associated regular perfect fluid solution 
that satisfies
\begin{equation}\label{sigass}
\sigma  < \frac{2}{\sqrt{g}} - 1
\end{equation}
for some constant $g \in [0,4)$, obeys the inequality $\Phi \geqslant g$ 
and consequently satisfies
\begin{equation}
\frac{2 M}{R} \leqslant \frac{8 - 2 g}{9-2 g}\:.
\end{equation}
\end{theorem}

\begin{remark}
In the special case $g = 0$, the assumption~\eqref{sigass} disappears. 
The theorem then reduces to the statement that every regular perfect fluid solution
satisfies the Buchdahl inequality. 
The Buchdahl inequality is thus naturally included as a special case of the theorem;
note in addition that the proof of the Buchdahl inequality becomes particularly simple
in our set-up.
\end{remark}

Considering the special case $g = 1$ in the theorem leads to an interesting corollary:

\begin{corollary}
Consider an equation of state $\rho = \rho(p)$ that is non-decreasing.
Every associated regular perfect fluid solution 
that satisfies the dominant energy condition (i.e., $\sigma = p/\rho < 1$)
obeys the inequality $\Phi \geqslant 1$ 
and consequently satisfies
\begin{equation}
\frac{2 M}{R} \leqslant \frac{6}{7}\:.
\end{equation}
\end{corollary}

In the subsequent section we give a proof of the theorem
which is based on the dynamical systems formulation of
the problem.

\section{Proof of the theorem}
\label{sec:dynamicalsystemsanalysis}

We begin by proving that, in the dynamical systems description, 
regular solutions satisfy $U \leqslant 3/4$ (where $U\rightarrow 3/4$
as $\lambda \rightarrow -\infty$), see also~\cite{Heinzle/Rohr/Uggla:2003}.
This statement is almost trivial, see the subsequent remark; 
however, we give a rather illustrative proof 
which already makes use of those 
methods that will turn out to be pivotal in the proper proof of
Theorem~\ref{thethm}.

\begin{remark}
In standard variables, the statement $U\leqslant 3/4$ reads
$\rho(r) \leqslant \bar{\rho}(r)$, where $\bar{\rho} = 3m/(4\pi r^3)$
is the ``average density''. This inequality is rather obvious, since
the density is outward decreasing, which is true provided that
the equation of state is non-decreasing.
\end{remark}

\begin{lemma}
Consider an equation of state $\rho = \rho(p)$ that is non-decreasing.
Then $U \leqslant 3/4$ holds for all regular perfect fluid solutions associated
with this equation of state.
\end{lemma}

\begin{proof}
Consider the surface $U = 3/4$ in the state space and the flow of
the system~\eqref{UVOmega} through this surface. 
We obtain $dU/d\lambda|_{U=3/4} =
-3\Upsilon(1+\sigma)(1+3\sigma)V/64\leqslant 0$, 
since $\Upsilon\geq 0$ when $d\rho/dp\geq 0$.
(One can say that the surface $U=3/4$ acts as a ``semi-permeable
membrane'' for the flow of the dynamical system, since solutions
can pass through this surface in only one way.)
Similarly, for a surface $U = U_0 > 3/4$ we obtain
$dU/d\lambda|_{U=U_0} < 0$.
Since, for regular solutions, $U\rightarrow 3/4$ as $\lambda\rightarrow -\infty$,
it follows that $U \rightarrow 3/4$ from below as $\lambda \rightarrow -\infty$ 
and that $U\leqslant 3/4$ for all $\lambda$.
\end{proof}

We now proceed to give the proof of \textbf{Theorem~\ref{thethm}}.
Note that (a variant of) the proof in the special case
$g= 0$ (the proper Buchdahl case) has been given 
already in~\cite{Heinzle/Rohr/Uggla:2003}. 
Here we extend these techniques.

\begin{proof}
Using that
\begin{equation}
\mathcal{M} =  
\frac{\sigma V}{1 - V + 2\sigma V}\,,\qquad
\mathcal{P} = 
\left(\frac{U}{1-U}\right)
\left(\frac{\sigma^2 V}{1 - V + 2\sigma V}\right)\:,
\end{equation}
the function $\Phi$, cf.~\eqref{Phidef}, 
can be regarded as a function on the state space.
In the state space description, regular solutions
emerge from the line of regular initial values 
$\{(U,V,\Omega)\,|\, U=3/4, V= 0, \Omega \in(0,1)\}$.
Evaluated in this limit, $\Phi$ becomes the function
\begin{equation}
\Phi\big|_{\{U=\frac{3}{4},V=0\}} = \frac{2 (2 + 3 \sigma)}{(1 +\sigma)^2}\:,
\end{equation}
whose values range in $[0,4)$. We obtain
\begin{equation}\label{sigmax}
\Phi\big|_{\{U=\frac{3}{4},V=0\}} = g = \frac{2 (2 + 3 \sigma)}{(1+\sigma)^2} \in [0,4) \quad\Leftrightarrow\quad
\sigma = \frac{3 - g + \sqrt{9 - 2 g}}{g} \in (0,\infty]\:.
\end{equation}
For each constant $g \in [ 0, 4)$,
the equation $\Phi = g$
defines a surface in the state space that 
divides the state space in two halves.
We are interested in the subset
\begin{equation}
\Phi > g\:,
\end{equation}
since this represents the Buchdahl type inequality we want to prove.
According to~\eqref{sigmax}, the subset $\Phi > g$ contains the 
line of regular initial values up to a certain value of $\Omega$, which
is given as the (smallest) value of $\Omega$ such that $\sigma <  \sigma_g$,
where $\sigma_g = (3 - g + \sqrt{9 - 2 g})/g$.
This means that, initially, i.e., for sufficiently small $\lambda$ (or, equivalently, for
sufficiently small $r$), regular solutions
with $\sigma < \sigma_g$ satisfy $\Phi > g$.

Let us consider a regular solution with 
\begin{equation}\label{sigcond}
\sigma <  \frac{2}{\sqrt{g}} - 1 
\end{equation}
and the function $\Phi$ along this solution, i.e., we
track $\Phi(\lambda) = \Phi\big(U(\lambda),V(\lambda),\Omega(\lambda)\big)$ for $\lambda \in (-\infty,\infty)$.
Obviously, $\Phi > g$ holds initially (i.e., for sufficiently small $\lambda$), 
since the r.h.~side of~\eqref{sigcond} is always less than $\sigma_g$.
In Lemma~\ref{dphiposlemma} we prove that the derivative of $\Phi$, when computed on the surface $\Phi =g$,
is positive under the assumption~\eqref{sigcond}, when we also use that $U\leqslant 3/4$; i.e.,
\begin{equation}\label{dPhi}
\frac{d\Phi}{d\lambda} \Big|_{\Phi = g} \geqslant 0\:.
\end{equation}
Therefore, since, first, $\Phi(\lambda) > g$ for sufficiently small $\lambda$ and,
second, $\Phi(\lambda)$ cannot pass through the $\Phi = g$ barrier, it follows
that
\begin{equation}
\Phi > g
\end{equation}
for all $\lambda \in \mathbb{R}$.
Equivalently, $\Phi > g$ for all $r \in [0,R)$ and $\Phi \geqslant g$ for all $r\in[0,R]$, 
and the theorem is established.
\end{proof}

It merely remains to prove the following lemma.

\begin{lemma}\label{dphiposlemma}
The derivative of $\Phi$ on the surface $\Phi =g$
is positive for all $U < 3/4$ 
if and only if 
\begin{equation}
\sigma <  \frac{2}{\sqrt{g}} - 1 \:.
\end{equation}
\end{lemma}

\begin{proof}
A straightforward computation based on the system~\eqref{UVOmega} yields
\begin{subequations}
\begin{align}
\frac{d\Phi}{d\lambda} \Big|_{\Phi = g} & = \frac{(3 - 4 U) (1 - U + \sigma U)}{3(1-U) +\sigma U} \: \sigma V\, (9 - 2 g)\: \varphi_g \:,
\intertext{where $\varphi_g$ is given by}
\varphi_g & = \frac{36 (1-U)^2 - g \,(3 (1-U) +\sigma U)^2}{18 (1-U) (2(1-U)+\sigma U) - g\, (3(1-U)+\sigma U)^2}\:.
\end{align}
\end{subequations}
Clearly, if $U < 3/4$,
\begin{equation}
\frac{d\Phi}{d\lambda} \Big|_{\Phi = g} > 0 \quad\text{if and only if}\quad
\varphi_g > 0\:.
\end{equation}
Since the denominator of $\varphi_g$ is equal to the numerator of $\varphi_g$ (which we call
$\hat{\varphi}_g$ in the following) plus a positive term, namely $18 (1-U) \sigma U$,
positivity of $\hat{\varphi}_g$ already implies positivity of $\varphi_g$.
Therefore, 
\begin{equation}
\frac{d\Phi}{d\lambda} \Big|_{\Phi = g} > 0 \quad\text{if}\quad
\hat{\varphi}_g = 36 (1-U)^2 - g \,(3 (1-U) +\sigma U)^2 > 0 \:.
\end{equation}
Positivity of $\hat{\varphi}_g$ is in turn governed by a simple linear condition,
\begin{equation}
\hat{\varphi}_g > 0 \quad\Leftrightarrow \quad
6(1- U) - \sqrt{g}\, (3 (1-U) + \sigma U) > 0 \:.
\end{equation}
The derivative of the l.h.~side w.r.t.\ $U$ is given by 
$-3 (2 -\sqrt{g}) - \sigma \sqrt{g}$, which is negative.
Therefore, the inequality is satisfied for all $U \leqslant 3/4$ if 
and only if it is satisfied for $U = 3/4$. Accordingly,
\begin{align}
\hat{\varphi}_g > 0  & \quad \Leftrightarrow \quad
\frac{3}{4} \left( 2 -\sqrt{g}\, (1+\sigma) \right) > 0 \\
&\quad \Leftrightarrow \quad
\sigma < \frac{2}{\sqrt{g}} - 1 \:,
\end{align}
which proves the claim of the lemma.
\end{proof}

\begin{remark}
The lemma shows that the surface $\Phi = g$ acts like a semi-permeable
membrane in the state space (provided that $U < 3/4$ and $\sigma <2/\sqrt{g} -1$).
Solutions that satisfy $\Phi > g$ cannot pass through the surface $\Phi = g$ and
are therefore confined to the subset $\Phi > g$ of the state space.
\end{remark}

\section{Conclusions and outlook}
\label{sec:conclusions}

In this paper we have proved a theorem that
formulates a straightforward and simple generalization of the Buchdahl inequality
for spherical models. Our considerations are based on
the assumption of perfect fluid matter that satisfies 
a non-decreasing equation of state $\rho(p)$;
however, it is clear that the theorem can also be applied
to collisionless matter (i.e., to the Vlasov-Einstein case),
provided that one restricts oneself to isotropic configurations.
We choose to not pursue the Vlasov-Einstein case further here;
instead we will provide a comprehensive discussion of
Vlasov-Einstein configurations (including some non-isotropic
cases) and an associated generalization of Theorem~\ref{thethm} 
in another publication.

In~\cite{Karageorgis/Stalker:2007}, assuming dominant energy,
an inequality is derived, namely inequality (4.3) of that paper, which 
gives a bound on $2 M/R$ by a number that is larger than $6/7$.
This inequality is claimed to be sharp; hence, in the light of
Theorem~\ref{thethm}, the solution with the maximal value of $2 M/R$
given by~\cite[(4.3)]{Karageorgis/Stalker:2007}
must violate at least one of our assumptions.

It is important to note that Theorem~\ref{thethm}
makes no statement about whether the obtained inequality is
sharp. (Trivially, it is sharp in the case $g=0$, which is
the original Buchdahl case, and in the case $g =4$.)
There might thus be room for improvement; whether this could be
achieved based on the methods used in the present paper, 
by the techniques used in~\cite{Karageorgis/Stalker:2007}, 
or a combination thereof, remains to be seen.



\end{document}